\documentclass[preprint]{aastex}

\def\gtorder{\mathrel{\raise.3ex\hbox{$>$}\mkern-14mu
                \lower0.6ex\hbox{$\sim$}}}
\def\ltorder{\mathrel{\raise.3ex\hbox{$<$}\mkern-14mu
                \lower0.6ex\hbox{$\sim$}}}
\newcommand{\hii}{H~{\small II} }

\shorttitle{A New Catalog of Compact \hii Regions}
\shortauthors{Giveon et al.}

\received{2005 January 17}
\begin{document}

\title{A New Radio Catalog of Compact \hii Regions in the Milky Way II. The 1.4 GHz Data}

\author{Giveon, U.}
\affil{Department of Physics, University of California, Davis, CA 95616;\\ giveon@physics.ucdavis.edu}
\author{Becker, R.H.}
\affil{Lawrence Livermore National Laboratory, Livermore, CA 94566;\\ bob@igpp.ucllnl.org}
\author{Helfand, D.J.}
\affil{Columbia Astrophysics Laboratory, 550 West 120th Street, New York, NY 10027;\\ djh@astro.columbia.edu}
\and
\author{White, R.L.}
\affil{Space Telescope Science Institute, 3700 San Martin Drive, Baltimore, MD 21218;\\ rlw@stsci.edu}

\begin{abstract}
We utilize a new VLA Galactic plane catalog at 1.4 GHz covering the first and second Galactic quadrants ($340^{\circ}\le l\le 120^{\circ}$, $|b|\le 0.8^{\circ}$ with $|b|\le 1.8^{\circ}$ for $350^{\circ}\le l\le 40^{\circ}$ and $|b|\le 2.5^{\circ}$ for $100^{\circ}\le l\le 105^{\circ}$) in conjunction with the MSX6C Galactic plane mid-infrared catalog to supplement and better understand our 5 GHz catalog (Giveon et al. 2005). A radio catalog of this region was first published by Zoonematkermani et al. (1990), but we have re-reduced the data with significantly improved calibration and mosaicing procedures, resulting in more than a tripling of the number of 1.4 GHz sources detected. Comparison of the new 1.4 GHz catalog and the MSX6C catalog resulted in a sample of 556 matches, out of which we estimate only 11 to be chance coincidences. Most of the matches show red MSX colors. The scale height of their Galactic latitude distribution is $24'-28'$ (FWHM) or $\sim 60-70$ pc (for a distance of 8.5 kpc), depending on longitude. The latitude distribution flattens out significantly above $l\approx 40^{\circ}$ and the number of matches drops off sharply.
\end{abstract}


\section{Introduction}
\label{intro}

Combining infrared and radio surveys provides a powerful approach to identifying specific classes of objects (i.e., \hii regions, ultra-compact [UC] \hii regions, planetary nebulae [PNe], and OH masers associated with star formation) based on the simultaneous presence of the cold molecular and dust phases and hot ionized gas (White, Becker \& Helfand 1991; Becker et al. 1994; Szymczak et al. 2002). In a recent paper (Giveon et al. 2005; hereafter Paper I), we have matched a 5 GHz VLA survey of the first quadrant of the Galactic plane with the MSX6C catalog (Egan, Price, \& Kramer 2003) to construct a large sample of compact \hii regions.

In this work we utilize a new VLA Galactic plane catalog at 1.4 GHz in conjunction with the MSX6C Galactic plane catalog to complement the 5 GHz catalog, which covers only a narrow $\ltorder 1^{\circ}$ strip along 52 degrees of the Galactic plane. The 1.4 GHz catalog covers a much larger area, both in latitude and longitude, and allows a better understanding of the spatial distribution of thermal sources.
In \S\ref{6cm} the new radio catalog is briefly introduced.
In \S\ref{iradio} we describe the matching of the 1.4 GHz radio catalog with the MSX6C catalog, and in \S\ref{properties} we study the spatial and infrared-color properties of the resulting subset of sources. In \S\ref{conc} we summarize our conclusions.

\section{The VLA Galactic Plane Survey at 1.4 GHz}
\label{6cm}

The original 1.4 GHz observations, described in detail in Zoonematkermani et al. (1990), cover the inner Galaxy ($340^{\circ}\le l\le 120^{\circ}$, $|b|\le 0.8^{\circ}$ with $|b|\le 1.8^{\circ}$ for $350^{\circ}\le l\le 40^{\circ}$ and $|b|\le 2.5^{\circ}$ for $100^{\circ}\le l\le 105^{\circ}$) with a resolution of $\sim 5''$ and sensitivity of $\sim 5-10$ mJy. Most of the original observations were taken in the VLA B configuration, with the southernmost fields observed in the A/B configuration. We applied improved phase calibration and reduction procedures, resulting in an increase in the number of sources detected by more than a factor of three from 1992 sources to 7051.
The principal improvement resulted from mosaicing areas of overlapping coverage which improved map sensitivity, particularly at the edges of the VLA primary beam. The new 1.4 GHz catalog, to be published separately (White, Becker, \& Helfand 2005\footnote{The 1.4 GHz and 5 GHz catalogs are available for download from the MAGPIS website at http://third.ucllnl.org/gps/.}), serves as the primary radio catalog for our comparison with the MSX6C catalog.

The improved reduction increased significantly the effective sensitivity of the 1.4 GHz catalog; it is now $>90\%$ complete for sources with $F_{1.4 GHz}\ge 10$ mJy, compared to 75\% completeness for 25 mJy sources in the earlier version. Adopting stellar fluxes from Sternberg, Hoffmann, \& Pauldrach (2003), we calculate that this allows us to detect at the far edge of the Galaxy (distance of 20 kpc) a B0 star with an ionization bounded nebula that is optically thin to free-free emission ($F_{1.4 GHz}=30$ mJy). However, the completeness level drops significantly even for the hottest O stars if the nebula is optically thick. For the extreme emission measures of UC \hii regions ($\sim 10^9$ pc$\cdot$cm$^{-6}$), our sample is only $\sim 50\%$ complete for detecting stars as late as O5. Expected radio and infrared flux densities for the hottest main sequence stars are listed in Table \ref{spec_type}. Infrared flux densities were calculated assuming 1\% of the total luminosity is reradiated in the MSX 8 $\mu$m band (based on the IRAS 12 $\mu$m band response stated by Wood \& Churchwell [1989] and on the MSX 8 $\mu$m band response function). The emission measure we used for the optically thick case corresponds in the most extreme case to a nebular gas density of $10^6$ cm$^{-3}$ and a nebular diameter of 0.003 pc (e.g., Mart\'{i}n-Hern\'{a}ndez, van der Hulst, \& Tielens 2003; Olmi \& Cesaroni 1999). For comparison, our 5 GHz survey (Paper I, Table 1) allows the detection of $>90\%$ of the most dense and compact cases, because the nebulae are relatively less optically thick for higher radio frequencies, and because the survey itself is deeper.

%%\clearpage
\begin{deluxetable}{ccccccc}
\tablecaption{Theoretical radio and infrared flux densities. \label{spec_type}}
\tablewidth{0pt}
\tablehead{
\colhead{Spectral} & \colhead{$T_{eff}$} & \colhead{$\log$ L}      & \colhead{$\log$ Q}      & \colhead{$F_{8\mu m}$} & \colhead{$F_{1.4GHz}$ ($\tau\ll 1$)} & \colhead{$F_{1.4GHz}$ ($\tau\gg 1$)} \\
\colhead{Type}     & \colhead{(K)}       & \colhead{($L_{\odot}$)} & \colhead{($\#/s^{-1}$)} & \colhead{(Jy)}     & \colhead{(mJy)}               & \colhead{(mJy)} \\
\colhead{(1)} & \colhead{(2)} & \colhead{(3)} & \colhead{(4)} & \colhead{(5)} & \colhead{(6)} & \colhead{(7)}}
\startdata
O3       &     51,000  &   6.3       &  49.9         & 2.2       &   2400      &   6.6      \\
O5       &     46,000  &   6.0       &  49.5         & 1.1       &    950      &   2.5      \\
O7       &     41,000  &   5.7       &  49.1         & 0.55      &    370      &   1.0      \\
O9       &     36,000  &   5.4       &  48.5         & 0.28      &     95      &   0.3      \\
O9.5     &     35,000  &   5.3       &  48.3         & 0.22      &     60      &   0.16     \\
B0       &     33,000  &   5.2       &  48.0         & 0.17      &     30      &   0.08     \\
B0.5     &     32,000  &   5.1       &  47.7         & 0.14      &     15      &   0.04     \\
\enddata
\tablecomments{Radio and infrared flux densities, for a distance of 20 kpc, based on model stellar spectra (Sternberg, Hoffmann, \& Pauldrach 2003). Columns list spectral types, effective temperatures, bolometric luminosities, ionizing photon fluxes $Q$, 8 $\mu$m flux densities, and 1.4 GHz flux densities for optically thin and optically thick ionization bounded nebulae. We have calculated the infrared flux densities assuming 1\% of the total luminosity is reradiated in the MSX 8 $\mu$m band (based on the IRAS 12 $\mu$m band response stated by Wood \& Churchwell [1989] and on the MSX 8 $\mu$m band response function). Optically thick 1.4 GHz flux densities were calculated using an emission measure of $\sim 10^9$ pc$\cdot$cm$^{-6}$.}
\end{deluxetable}
%%\clearpage

\section{Radio-Infrared Matching}
\label{iradio}

The MSX6C archival data (version 2.3, Egan, Price, \& Kramer 2003) comprises 
431,711 sources observed in four bands -- A, C, D, and E -- with central wavelengths 8.3, 12.1, 14.7, and 21.3 $\mu$m, respectively. Hereafter we will refer to the MSX bands by their central wavelengths -- 8, 12, 14, and 21 $\mu$m. The MSX6C survey covers the entire Galaxy for latitudes $|b|\le 6^{\circ}$ with a resolution of $18.3''$. The sensitivity ranges from 0.1 Jy at 8 $\mu$m to 6 Jy at 21 $\mu$m. The MSX6C coverage is not uniform over the Galactic plane for $|b|\le 6^{\circ}$ owing to varying scan overlap resulting from a change in the scan rate at the end of the mission (Egan, Price, \& Kramer 2003).

We match the 1.4 GHz catalog and the MSX6C catalog by considering two criteria: $\alpha$ - a conservative matching radius of $12''$, and $\beta$ - an additional annulus with $12''<r\le 25''$; adding this larger annulus still yields real matches. We estimate the number of false matches by comparing the MSX6C catalog to false radio catalogs with the same spatial distribution by shifting the original catalog by $\pm 10'$ and $\pm 20'$ in Galactic longitude. The results from the matches for each criterion are given in Table \ref{proba1}: number of total matches (column 3), and number of false matches (column 4) from the shifted radio maps. In the real catalog, however, the false match rate will be lower than in the shifted catalog since true matches cannot also be false matches. Hence, we correct the false match rate for the fraction of true matches by a factor of $1-F/N$, where $F$ is the number of false matches and $N$ is the total number of sources in the radio catalog. The true match rates that correspond to these corrected false rates are listed in column 5.

%\clearpage

\begin{deluxetable}{ccccc}
\tablecaption{Matching Criteria and Results.\label{proba1}}
\tablewidth{0pt}
\tablehead{
 \colhead{} & \colhead{Matching Radius} &  \colhead{} &  \colhead{} & \colhead{True Matches} \\
 \colhead{Matching Criterion} & \colhead{(arcsec)}        & \colhead{Total Matches} &  \colhead{False Matches} &  \colhead{(\%)} \\
\colhead{(1)} & \colhead{(2)} & \colhead{(3)} & \colhead{(4)} & \colhead{(5)}}
\startdata
$\alpha$ &$12''$      & 686 & 117 & 84 \\
$\beta$  &$12''-25''$ & 471 & 372 & 22 \\
\enddata
\end{deluxetable}
%\clearpage

We have used the information on the MSX6C bands in which a given source was detected to refine the probability calculation in Table \ref{proba1} (see Paper I). In principle, the distribution of false matches should mimic the MSX6C band distribution, while MSX6C sources with 1.4 GHz emission may have quite a different distribution of band detections. Table \ref{proba2} lists the results of these probability calculations. We have counted the number of detections in each band combination for the MSX6C sources in the Galactic region roughly defined by the radio catalog. Column 2 lists these numbers, while column 3 lists their percentage of the total number of sources. We have calculated the expected false rates (columns 4 \& 5) for these numbers according to the values given in Table \ref{proba1}.
The actual matches found in each band combination are given in columns 6 \& 7, while the final match reliabilities are given in columns 8 \& 9.

%\clearpage
\begin{deluxetable}{lcccccccc}
\tabletypesize{\scriptsize}
\tablecaption{Matching Reliability Using Band Detection Information. \label{proba2}}
\tablewidth{0pt}
\tablehead{
\colhead{} & \colhead{} & \colhead{Portion of MSX Sources} & \multicolumn{2}{c}{Expected False Matches} & \multicolumn{2}{c}{Actual Total Matches} & \multicolumn{2}{c}{Reliability} \\
\colhead{MSX Bands} &  \colhead{Number of MSX Sources\tablenotemark{a}} & \colhead{(\%)} & \colhead{$\alpha$} & \colhead{$\beta$} & \colhead{$\alpha$} & \colhead{$\beta$} & \colhead{$\alpha$} & \colhead{$\beta$} \\
\colhead{(1)} & \colhead{(2)} & \colhead{(3)} & \colhead{(4)} & \colhead{(5)} & \colhead{(6)} & \colhead{(7)} & \colhead{(8)} & \colhead{(9)}}
\startdata
8 only\tablenotemark{b} & 98868 & 59.2  & 69    & 220   &  58 & 180 & 0.00 & 0.00 \\
12 only    &     4 &  0    &  0    &   0   &   0 &   0 & 0.00 & 0.00 \\
14 only    &     0 &  0    &  0    &   0   &   0 &   0 & 0.00 & 0.00 \\
21 only    &    51 &  0.03 &  0    &   0.1 &   0 &   0 & 0.00 & 0.00 \\
8-12\tablenotemark{b}   & 13976 &  8.4  &  9.8  &  31   &   5 &  25 & 0.00 & 0.00 \\
8-14\tablenotemark{b}   &  9049 &  5.4  &  6.3  &  20   &  15 &  13 & 0.61 & 0.00 \\
8-21       &  3465 &  2.1  &  2.5  &   7.8 &   4 &   8 & 0.43 & 0.04 \\
12-14      &     2 &  0    &  0    &   0   &   0 &   0 & 0.00 & 0.00 \\
12-21      &     7 &  0    &  0    &   0   &   0 &   0 & 0.00 & 0.00 \\
14-21      &   178 &  0.11 &  0.13 &   0.4 &   0 &   0 & 0.00 & 0.00 \\
8-12-14\tablenotemark{b}& 20597 & 12.3  & 14.4  &  46   &  16 &  33 & 0.83 & 0.00 \\
8-12-21\tablenotemark{b}&  1102 &  0.66 &  0.8  &   2.5 &   2 &   1 & 0.63 & 0.00 \\
8-14-21    &  1131 &  0.68 &  0.8  &   2.5 &  42 &   6 & 0.98 & 0.59 \\
12-14-21   &   193 &  0.12 &  0.14 &   0.4 &   4 &   1 & 0.97 & 0.61 \\
8-12-14-21 & 18318 & 11.0  & 12.9  &  41   & 540 & 204 & 0.98 & 0.80 \\
\enddata
\tablenotetext{a}{Taken from a limited area with coordinates ranges approximately the same as the 1.4 GHz catalog.}
\tablenotetext{b}{The expected false rate in these band combinations is significantly overestimated (mainly for the $\beta$ criterion). The large discrepancy between expected and actual false rates is a result of these sources being dominated by stars, which have a flat latitude distribution in the survey area, while the distribution of the radio sources peaks at the Galactic plane.}
\end{deluxetable}
%\clearpage

We find that 90\% of the $\alpha$ criterion matches are $\ge50\%$ reliable, and 45\% of the $\beta$ criterion matches are $\ge 50\%$ reliable. A total of 586 matches have reliability $\ge 97\%$; only 12 are expected to be false. This is a major improvement over the color-independent values in Table \ref{proba1}. The poor sensitivity of the 1.4 GHz survey compared to the 5 GHz survey is reflected in the smaller number of matches despite the larger coverage area (see Paper I).
For point sources (radio size $\le 5''$) with high-reliability matches ($\ge 97\%$) we find that the astrometry of the MSX6C catalog and the VLA 1.4 GHz is consistent within errors of $3''$.

Out of the 586 matches with $\ge 97\%$ reliability, 4 are multiple MSX6C matches -- cases where one radio source has two or three MSX6C counterparts within the matching radius -- and 199 are multiple radio matches -- cases where one MSX6C source has two or more radio counterparts. For the latter case, since the reliabilities are comparable, the multiplicity implies two or more peaks in the radio emission which are unresolved in the infrared by the coarser spatial resolution of the MSX6C catalog.

From the matching subset of 582 single radio sources, 266 fall within the area covered by the 5 GHz survey ($350^{\circ}\le l\le 42^{\circ}$, $|b|\le 0.4^{\circ}$). In Paper I we mention that there are 228 MSX6C-5 GHz matches that also have entries in the 1.4 GHz catalog. Thus, there are 38 MSX6C-1.4 GHz matches that are not detected at 5 GHz. From these, 21 entries cannot be explained by multiple 1.4 GHz-5 GHz matches or choice of different 1.4 GHz-5 GHz matching radii. It turns out that 5 of these 21 sources are located in fields with degraded sensitivity at 5 GHz (out of which 3 are known \hii regions - one source has in fact a 5 GHz counterpart that was not included in the catalog - 12.68430-0.18242). Using the rms noise levels from the 5 GHz fields of the other 16 sources gives upper limits on their radio spectral slopes, and all of them show values $\alpha\le -1.1$ ($F_{\nu}\propto\nu^{\alpha}$). This implies that, in spite the fact that they are high-reliability infrared emitters, they are probably non-thermal sources. Indeed, 13 out of the 16 were previously identified as OH masers (by e.g., Becker, White, \& Proctor 1992; Blommaert, van Langevelde, \& Michiels 1994; Sevenster et al. 1997).
There are in total 23 OH masers from Becker, White, \& Proctor (1992) that are included in our matching subset. Together with the 3 unidentified non-thermal sources they make 26 sources that we exclude from the following analysis. We include only the remaining 556 single radio sources that have high reliability ($\ge 97\%$) MSX6C matches, of which only 11 are estimated to be false matches. The total number of single sources with reliability $\ge 50\%$ is 746, with 49 estimated to be false matches.

Table \ref{source_list} lists the 556 high-reliability sources and an additional 190 lower-reliability sources. For each source we provide the Galactic coordinates; 1.4 GHz peak flux density in millijanskys/beam; 1.4 GHz integrated flux density in millijanskys; deconvolved major and minor axes in arcseconds; 5 GHz peak flux density in millijanskys/beam when a counterpart exists; 5 GHz integrated flux density in millijanskys when a counterpart exists; distance from the MSX6C source in arcseconds; name of the MSX6C match; 8, 12, 14, and 21 $\mu$m flux densities of the MSX6C match in janskys; and the reliability of the match. 

%\clearpage
\begin{deluxetable}{crrrrrcrrrrc}
\tabletypesize{\tiny}
\tablecaption{High Reliability MSX6C-1.4 GHz Matches. \label{source_list}}
\tablewidth{0pt}
\tablehead{
\colhead{} & \colhead{$F_p$} & \colhead{$F_i$} & \colhead{Major} & \colhead{Minor} & \colhead{Separation} & \colhead{} & \colhead{$F_8$} & \colhead{$F_{12}$} & \colhead{$F_{14}$} & \colhead{$F_{21}$} & \\
\colhead{Name} & \colhead{(mJy/beam)} & \colhead{(mJy)} &\colhead{(arcsec)} & \colhead{(arcsec)} & \colhead{(arcsec)} & \colhead{MSX6C Name} &  \colhead{(Jy)} &  \colhead{(Jy)} & \colhead{(Jy)} & \colhead{(Jy)} & \colhead{Reliability} \\
\colhead{(1)} & \colhead{(2)} & \colhead{(3)} & \colhead{(4)} & \colhead{(5)} & \colhead{(6)} & \colhead{(7)} & \colhead{(8)} & \colhead{(9)} & \colhead{(10)} & \colhead{(11)} & \colhead{(12)}}
\startdata
340.25011-0.04490 &    62.1 &    95.5 &  5.25 &  2.57 &  5.2 & G340.2490-00.0460 &    2.50 &    5.64 &   11.44 &   63.60 & 0.98 \\
340.05287-0.23133 &    19.8 &   408.8 & 33.08 & 16.57 &  8.1 & G340.0511-00.2334 &    6.31 &   19.88 &   26.35 &   98.27 & 0.98 \\
340.05348-0.23365 &    43.8 &   104.7 &  7.68 &  4.58 &  4.6 & G340.0511-00.2334 &    6.31 &   19.88 &   26.35 &   98.27 & 0.98 \\
340.05777-0.24269 &   136.9 &   215.5 &  4.61 &  3.41 &  7.9 & G340.0543-00.2437 &    2.09 &    3.57 &    7.59 &   58.33 & 0.98 \\
339.98117-0.53796 &    14.9 &    31.7 & 11.03 &  2.02 &  4.1 & G339.9797-00.5391 &    1.13 &    2.71 &    5.83 &   21.60 & 0.98 \\
340.25021-0.37077 &    33.6 &    52.5 &  6.17 &  2.16 &  6.5 & G340.2480-00.3725 &    4.03 &    9.09 &   13.35 &   54.75 & 0.98 \\
341.28565 0.11866 &    13.7 &    12.5 &  3.29 &  0.00 &  3.3 & G341.2847+00.1179 &    6.89 &   12.07 &   17.06 &   14.50 & 0.98 \\
341.21114-0.23044 &    15.2 &   475.9 & 45.23 & 18.68 & 10.0 & G341.2105-00.2325 &    6.11 &   14.36 &   16.48 &   39.99 & 0.98 \\
342.06157 0.42123 &    85.2 &   426.0 & 13.22 &  8.33 &  6.4 & G342.0610+00.4200 &    3.53 &   15.60 &   24.57 &   57.04 & 0.98 \\
342.06350 0.41864 &    23.8 &   149.3 & 14.32 &  9.94 &  9.7 & G342.0610+00.4200 &    3.53 &   15.60 &   24.57 &   57.04 & 0.98 \\
341.70309 0.05271 &    86.1 &   153.8 &  5.33 &  3.93 &  7.2 & G341.7018+00.0509 &    1.86 &    2.70 &    2.55 &    5.29 & 0.98 \\
342.22773-0.37973 &    91.2 &   133.7 &  5.46 &  2.26 &  0.9 & G342.2263-00.3801 &    0.71 &    1.01 &    6.24 &   13.70 & 0.98 \\
343.12918-0.05958 &     6.7 &     8.2 &  5.17 &  0.00 & 11.0 & G343.1261-00.0623 &    0.49 &   -0.55 &    0.52 &   15.89 & 0.98 \\
343.47670-0.02801 &     9.4 &   596.3 & 74.08 & 23.18 & 11.1 & G343.4766-00.0300 &    3.52 &   11.29 &   20.23 &   51.10 & 0.98 \\
343.47858-0.02743 &    13.1 &   617.6 & 58.34 & 21.84 &  9.9 & G343.4766-00.0300 &    3.52 &   11.29 &   20.23 &   51.10 & 0.98 \\
343.48095-0.02913 &    12.7 &   261.2 & 30.54 & 17.50 & 10.9 & G343.4766-00.0300 &    3.52 &   11.29 &   20.23 &   51.10 & 0.98 \\
343.50414-0.01303 &    95.2 &   222.5 &  6.58 &  5.14 &  5.2 & G343.5024-00.0145 &   14.43 &   25.93 &   28.64 &  130.72 & 0.98 \\
343.91264 0.11180 &     6.9 &    40.8 & 17.09 &  7.92 &  7.2 & G343.9101+00.1099 &    0.69 &    1.34 &    0.67 &    1.62 & 0.98 \\
344.42720 0.04701 &   273.4 &   355.3 &  4.16 &  0.00 &  6.9 & G344.4257+00.0451 &   22.72 &   78.40 &  125.36 &  445.02 & 0.98 \\
344.42538 0.04393 &    54.3 &   216.9 & 15.96 &  0.00 &  8.6 & G344.4257+00.0451 &   22.72 &   78.40 &  125.36 &  445.02 & 0.98 \\
344.42820 0.04611 &   184.7 &  1149.8 & 14.89 &  9.78 &  4.2 & G344.4257+00.0451 &   22.72 &   78.40 &  125.36 &  445.02 & 0.98 \\
344.22270-0.59307 &   195.7 &   326.8 &  4.71 &  3.81 &  8.5 & G344.2207-00.5953 &   13.67 &   26.98 &   29.94 &  130.65 & 0.98 \\
344.22364-0.59404 &   101.6 &   542.9 & 11.33 & 10.48 &  6.4 & G344.2207-00.5953 &   13.67 &   26.98 &   29.94 &  130.65 & 0.98 \\
345.49742 0.35383 &    15.2 &    67.1 & 25.10 &  2.43 &  7.2 & G345.4970+00.3525 &    9.72 &   25.74 &   34.14 &  133.95 & 0.98 \\
345.49910 0.35312 &    51.6 &   160.7 & 12.64 &  4.19 &  2.2 & G345.4970+00.3525 &    9.72 &   25.74 &   34.14 &  133.95 & 0.98 \\
345.49007 0.31630 &   316.5 &   993.6 &  8.13 &  7.11 &  5.1 & G345.4881+00.3148 &   13.52 &   39.63 &   66.60 &  256.20 & 0.98 \\
345.48089 0.14077 &    30.5 &   100.8 &  9.46 &  6.76 &  6.7 & G345.4800+00.1392 &    0.16 &    1.04 &    3.11 &    8.16 & 0.98 \\
345.52898-0.04991 &    27.8 &    66.1 &  7.81 &  3.86 &  5.9 & G345.5285-00.0508 &    2.23 &    5.37 &    5.41 &   10.50 & 0.98 \\
345.65062 0.01059 &   310.3 &  1440.8 & 10.81 &  9.05 &  8.9 & G345.6495+00.0084 &    6.26 &   18.17 &   28.79 &   80.01 & 0.98 \\
345.54889-0.07911 &    31.8 &    38.3 &  3.51 &  1.44 &  3.1 & G345.5472-00.0801 &    0.44 &    0.69 &    0.46 &    3.76 & 0.98 \\
345.80561 0.04904 &    21.2 &    32.4 &  6.72 &  1.63 &  3.2 & G345.8044+00.0481 &    1.34 &    2.09 &    2.25 &    9.83 & 0.98 \\
345.69982-0.08861 &    12.5 &    12.0 &  2.47 &  0.00 &  2.8 & G345.6985-00.0894 &    0.65 &    2.96 &    6.86 &   20.00 & 0.98 \\
346.05893-0.02330 &     8.6 &    49.5 & 17.91 &  7.39 &  2.9 & G346.0578-00.0229 &    0.78 &    3.06 &    4.00 &   12.01 & 0.98 \\
346.07844-0.05437 &    26.4 &    70.6 &  9.29 &  4.98 &  7.7 & G346.0774-00.0562 &    4.43 &    8.94 &    8.56 &   19.41 & 0.98 \\
346.52374 0.08578 &   120.2 &   198.0 &  7.62 &  0.53 &  9.6 & G346.5235+00.0839 &    2.18 &    6.26 &   10.01 &   27.00 & 0.98 \\
346.52455 0.08369 &    76.6 &   379.1 & 13.36 &  7.98 &  2.8 & G346.5235+00.0839 &    2.18 &    6.26 &   10.01 &   27.00 & 0.98 \\
347.60180 0.24459 &    18.6 &    40.4 &  7.12 &  4.59 &  1.6 & G347.5998+00.2442 &    1.73 &    4.30 &    5.06 &   16.57 & 0.98 \\
347.61617 0.15215 &    48.6 &   151.4 & 11.68 &  3.54 &  6.8 & G347.6162+00.1517 &    7.64 &   17.20 &   17.02 &   46.75 & 0.98 \\
347.62040 0.15362 &    17.4 &   112.8 & 18.24 &  6.98 & 11.0 & G347.6162+00.1517 &    7.64 &   17.20 &   17.02 &   46.75 & 0.98 \\
347.61729 0.15088 &    33.5 &   133.6 & 15.24 &  3.01 &  4.7 & G347.6162+00.1517 &    7.64 &   17.20 &   17.02 &   46.75 & 0.98 \\
347.61881 0.15089 &     9.9 &   218.6 & 31.27 & 18.49 &  6.8 & G347.6162+00.1517 &    7.64 &   17.20 &   17.02 &   46.75 & 0.98 \\
347.61916 0.14769 &     8.3 &    62.9 & 20.78 &  8.55 &  1.8 & G347.6177+00.1470 &    3.09 &   17.19 &   25.01 &   64.11 & 0.98 \\
347.94643 0.35315 &     8.8 &     7.0 &  3.31 &  0.00 &  1.8 & G347.9450+00.3524 &    9.13 &   12.30 &   19.49 &   18.05 & 0.98 \\
347.89808 0.04398 &     6.9 &   166.0 & 56.81 & 10.95 & 10.8 & G347.8962+00.0411 &    3.69 &   10.14 &   11.65 &   32.22 & 0.98 \\
347.89727 0.04272 &    17.7 &   301.1 & 36.80 & 12.00 &  6.4 & G347.8962+00.0411 &    3.69 &   10.14 &   11.65 &   32.22 & 0.98 \\
347.90555 0.04839 &    19.4 &   166.6 & 20.67 &  9.87 &  6.8 & G347.9023+00.0481 &    7.02 &   20.83 &   25.10 &   72.19 & 0.98 \\
347.87203 0.01462 &    77.4 &    96.3 &  3.38 &  2.00 &  1.1 & G347.8707+00.0146 &    0.40 &    0.79 &    0.63 &    8.22 & 0.98 \\
349.51042 1.05638 &   751.2 &  1617.3 &  7.00 &  3.40 &  2.5 & G349.5090+01.0555 &   12.93 &   22.75 &   67.91 &  211.96 & 0.98 \\
347.96807-0.43097 &    41.3 &    73.5 &  4.63 &  4.37 &  6.7 & G347.9676-00.4321 &    3.37 &    3.27 &    2.77 &   17.41 & 0.98 \\
347.97110-0.43050 &     6.6 &    19.0 &  9.77 &  5.35 &  8.3 & G347.9676-00.4321 &    3.37 &    3.27 &    2.77 &   17.41 & 0.98 \\
\enddata
\tablecomments{Table \ref{source_list} is presented in its entirety in the electronic edition of the journal. A portion is shown here for guidance regarding its form and content.}
\end{deluxetable}
%\clearpage

\section{Properties of the Matching Subset}
\label{properties}

\subsection{Infrared Colors}
\label{rad_ir}

\begin{figure}
\plotone{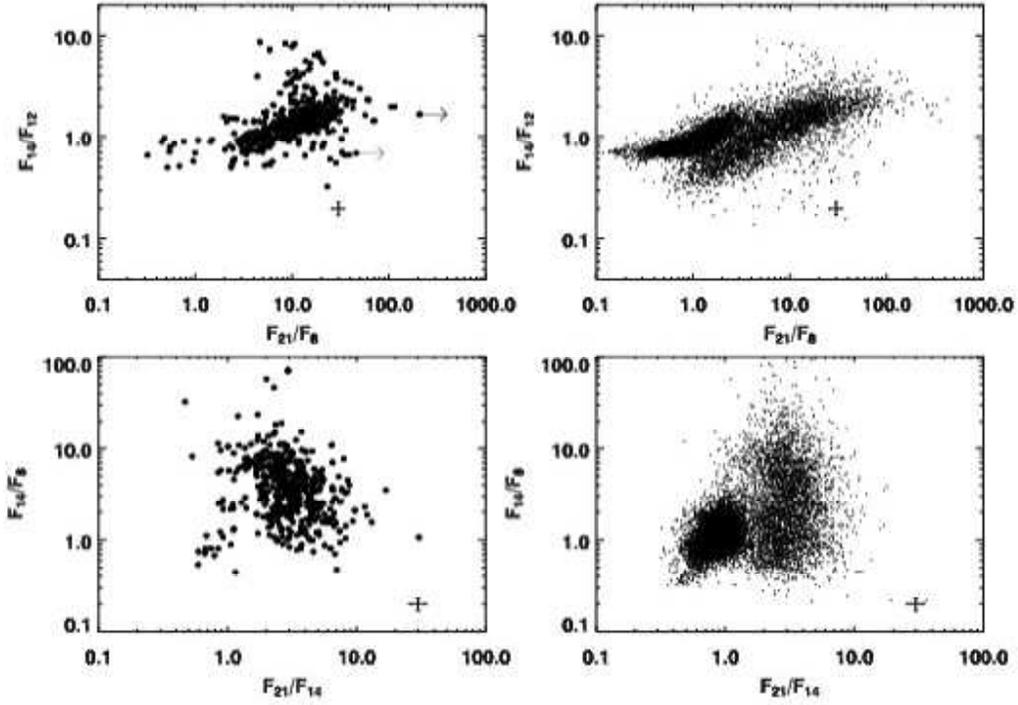}
\caption{Infrared color-color diagrams for the MSX6C-1.4 GHz matches (left panels) and for all the high-quality (Lumsden et al. 2002) MSX6C sources in the area of our radio survey (right panels). Representative error bars are plotted. Clearly, radio sources with infrared counterparts belong to the red MSX population.}
\label{radio_cc}
\end{figure}

We now investigate the infrared colors of the 556 radio sources with highly reliable MSX6C counterparts. Lumsden et al. (2002) show with an earlier version of the MSX Galactic plane catalog (MSX5C) that there are two underlying populations distinguished by their infrared colors. One population shows 'red' infrared colors, and is nebular in nature -- \hii regions, PNe, methanol masers without radio emission, and young stellar objects. Lumsden et al. have shown that these objects are heavily embedded sources, in which the infrared colors are determined mainly by the opacity and temperature of the dust surrounding them and not by the direct photospheric contribution of the central exciting star. The other population shows bluer colors, and consists mainly of evolved stars. Lumsden et al. based this result on detections of previously classified sources. In the left panels of Figure \ref{radio_cc} we plot two color-color diagrams for our sample of the MSX6C-1.4 GHz matches following Lumsden et al. For comparison, we plot
in the right panels all the MSX6C sources with high-quality flags detected in the corresponding bands in the area covered by the radio survey (quality flag criterion as in Lumsden et al.). This plot is slightly different from the parallel plot in Lumsden et al. which includes only sources in a narrow longitude range ($20^{\circ}\le l\le 30^{\circ}$), and in a broader latitude range ($|b|\le 6^{\circ}$).

%\clearpage
%\clearpage

Matching the 1.4 GHz and MSX6C catalogs filters out the majority blue population, leaving mostly red-population sources. For example, there are very few MSX6C-1.4 GHz matches with both $F_{21}/F_8 <2$ and $F_{14}/F_{12} >0.7$ (upper panels) or with $F_{21}/F_{14} <1.5$ (lower panels).
This result implies that our subset of matching sources consists mostly of \hii regions and PNe, since these two populations are the dominant radio emitters in the Galaxy. However, we cannot distinguish the infrared colors of these two populations.
From our subset of 556 sources, 47 were identified as PNe by White, Becker, \& Helfand (1991) and Becker et al. (1994) by matching the older versions of the 1.4 GHz and the 5 GHz maps with the IRAS catalog, and using the IRAS color criteria. Of these 47, 21 were ``candidates'' because only limits on their infrared flux densities were available. By considering the increase in source detection in the present version of the 1.4 GHz maps and assuming a constant PNe fraction, we estimate that there are $<164$ PNe in our sample ($<29\%$). This fraction of the matching subset is much larger than the $11\%$ stated in Paper I for the MSX6C-5 GHz matching, owing to the fact that the 1.4 GHz map covers a much broader area around $b=0^{\circ}$, including an increasing fraction of PNe whose distribution is more latitude-independent.
In the following sections we show results that still support the argument that the matching subset is dominated by \hii regions.

\begin{figure}
\plotone{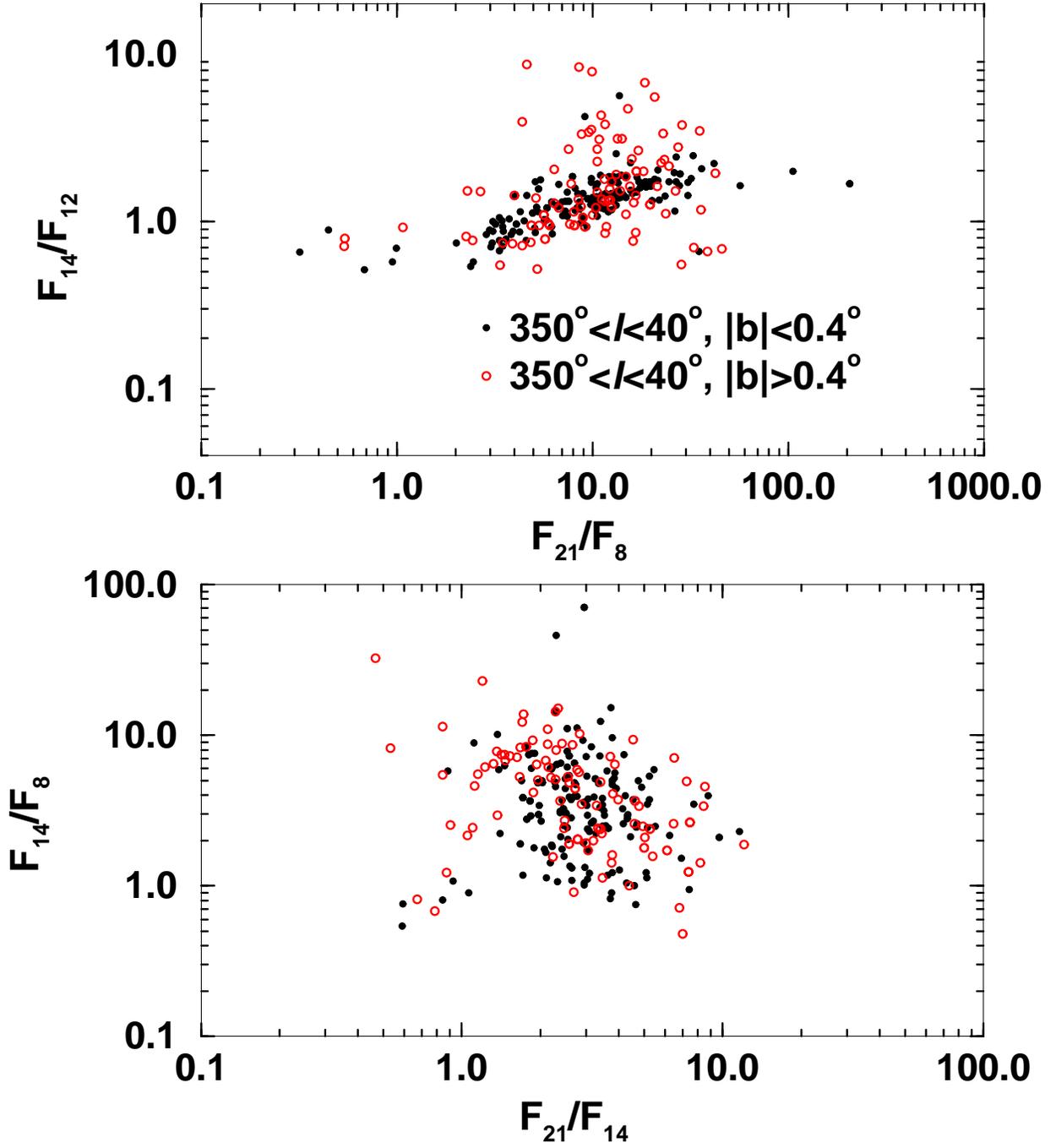}
\caption{Infrared color-color diagrams for the MSX6C-1.4 GHz matches in the area $350^{\circ}<l<40^{\circ}$ divided into lower-latitude (filled circles) and higher-latitude (open circles) bins.}
\label{lat_cc}
\end{figure}

The broader latitude strip along $350^{\circ}<l<40^{\circ}$ allows us to investigate whether color varies with latitude. Figure \ref{lat_cc} shows the MSX6C-1.4 GHz matches in that area divided into lower-latitude ($b\le 0.4^{\circ}$) and higher-latitude ($0.4^{\circ}<b<1.8^{\circ}$) bins.

%\clearpage
%\clearpage

The average flux ratios of the two latitude bins are consistent with each other, except for $F_{14}/F_{12}$: $1.38\pm 0.04$ for the lower-latitude bin and $1.87\pm 0.15$ for the higher-latitude bin. This may indicate a difference in the ratio of the [NeII]12.8$\mu$m and the [NeIII]15.6$\mu$m forbidden lines within the 12$\mu$m and the 14$\mu$m bands, respectively, which is proportional to the relative nebular ionization. Hence, higher $F_{14}/F_{12}$ on average, means higher relative ionization.
This flux ratio has smaller spread for lower-latitude sources: $\sigma=3\%$ compared to 8\% for the higher-latitude sources. This may reflect differences in the source populations: the youngest objects (mainly compact \hii regions), which are completely embedded in their natal dust cocoons, are more confined to the Galactic plane and have a tighter relation between their mid-infrared colors which mainly reflects the dust surrounding them; higher latitude sources may be a mixture of populations, including older \hii regions (and PNe), which already have part of their ionized gas unobscured with greater contributions of emission lines to the bands' fluxes. A spread in the relative strengths of these lines, reflecting a spread in nebular ionization, will also spread out the color-color distribution.

Previous identifications of some of the sources support this picture. Using the SIMBAD database\footnote{http://simbad.u-strasbg.fr/} with a matching radius of $10''$, we find that 145 sources out of the subset of 556 are identified as \hii regions, while 35 are identified as PNe. For the lower-latitude sources, we find that 90\% of the identifications are \hii regions and only 10\% are PNe, while for the higher-latitude sources, the fraction of the \hii regions goes down to 60\%, with PNe contributing 40\%.

\subsection{Spatial Distributions}
\label{rad_spat}

\begin{figure}
\plotone{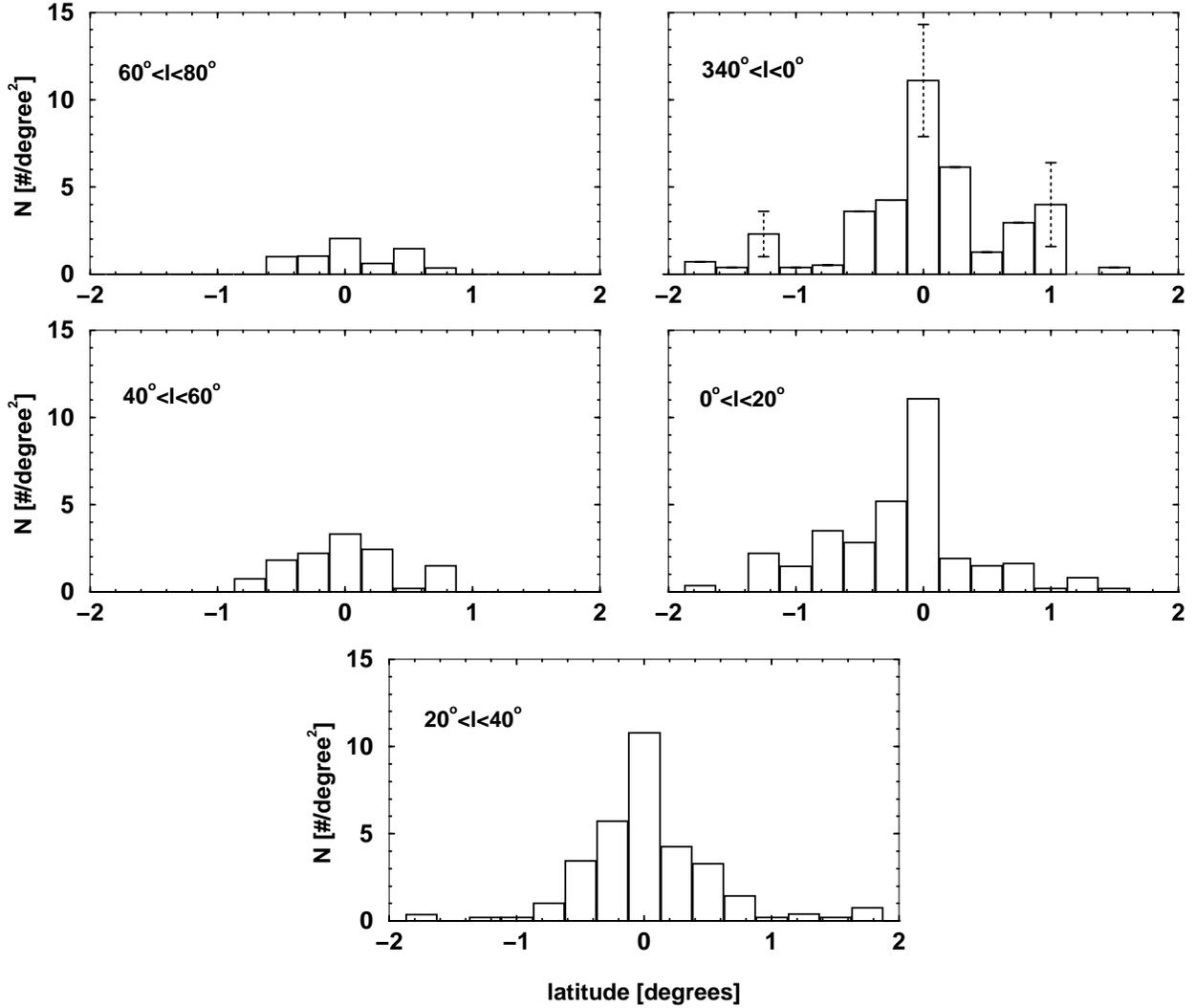}
\caption{Latitude distributions of the MSX6C-1.4 GHz matches in five consecutive longitude bins going from the top right panel through the bottom to the top left. The last two histograms flatten out, reflecting the decrease in Population I source detection for $l\gtorder 40^{\circ}$. Table \ref{bdist} expresses that behavior numerically. Figure \ref{coords_lb_20cm} shows how the average latitude varies with longitude in a two-dimensional plot. Representative error bars are shown in the top right panel.}
\label{coords_b_20cm}
\end{figure}

\begin{figure}
\plotone{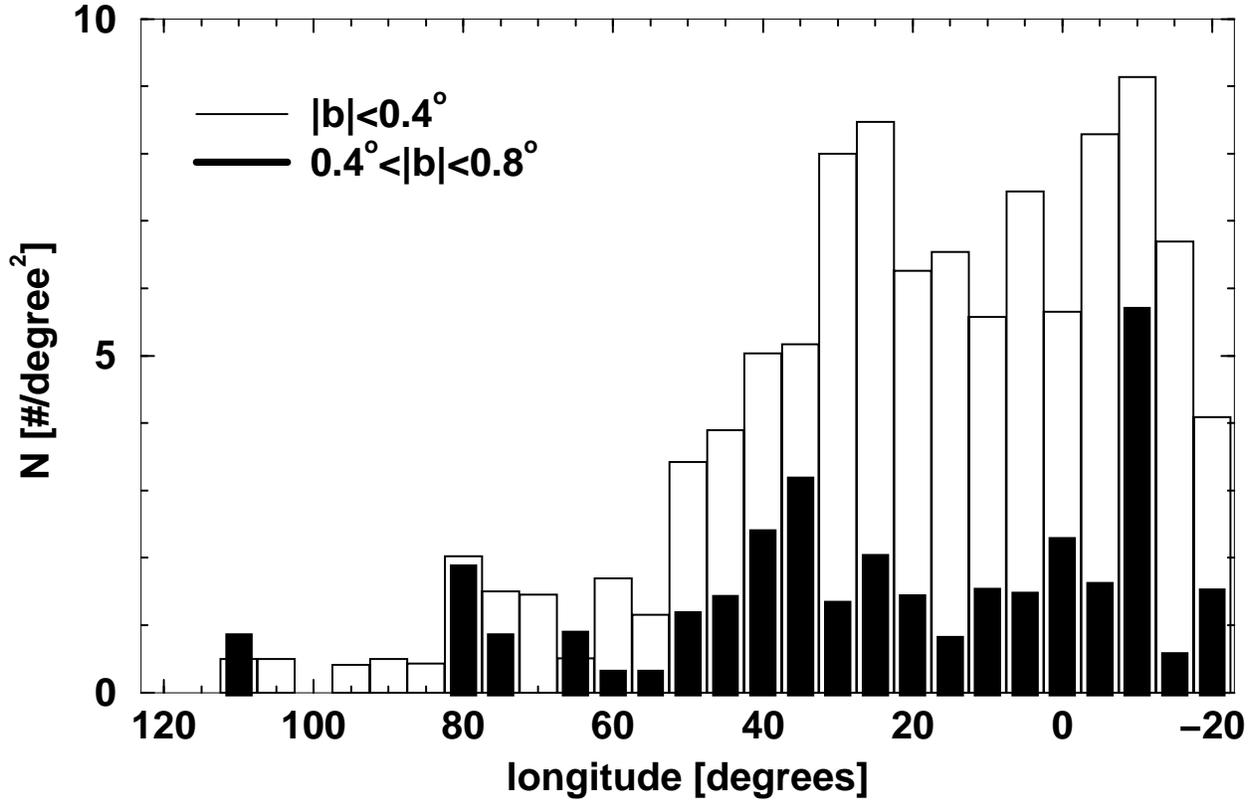}
\caption{Longitude distributions of the MSX6C-1.4 GHz matches with $|b|\le 0.4^{\circ}$ (open bars) and  $0.4^{\circ}<|b|\le 0.8^{\circ}$ (filled bars) for which the coverage is nearly uniform. For lower-latitude sources, the number of matching sources drops off significantly at $l\approx 40^{\circ}$ and almost completely vanishes $l\gtorder 80^{\circ}$, while for higher-latitude sources, the longitude distribution is much flatter.}
\label{coords_l_20cm}
\end{figure}

In Figures \ref{coords_b_20cm} and \ref{coords_l_20cm} we show the distribution in Galactic coordinates of the 556 radio sources with MSX6C counterparts. The latitude and longitude histograms are normalized using our survey coverage map (White, Becker, \& Helfand 2005) to correct for uneven coverage due to varying noise levels in the maps. The coverage was summed up as a function of flux density, and each flux density bin was assigned the corresponding coverage area (assuming that the flux density distribution is the same over the entire area covered). The raw latitude and longitude bins were then divided by the corresponding coverage area bin, with the underlying assumption that latitude and longitude are independent of each other in each bin.

In Figure \ref{coords_b_20cm} we show latitude histograms for five longitude bins, since there is no sense in integrating the latitude distribution over the entire longitude range, in which different parts may have completely different spatial properties.

%\clearpage
%\clearpage

The 1.4 GHz sources detected in the infrared show a narrow latitude distribution along the Galactic plane, with a FWHM of only $24'-28'$, depending on longitude, up to $l\approx 40^{\circ}$ (see Table \ref{bdist}). For a distance of 8.5 kpc to the Galactic center, this corresponds to $\sim 60-70$ pc. Even in its narrowest part, the width of the latitude distribution of the matching subset is larger than that of the 5 GHz catalog ($16'$; Paper I), probably because the 1.4 GHz has poorer sensitivity, increasing the fraction of closer B stars, which have a broader distribution relative to the more distant, faint O stars detectable at 5 GHz.
Table \ref{bdist} lists the means, widths, and peak heights of the latitude distributions as a function of longitude.

%\clearpage
\begin{deluxetable}{cccc}
\tablecaption{Variation in Latitude Distribution as a Function of Longitude. \label{bdist}}
\tablewidth{0pt}
\tablehead{
\colhead{Longitude Range} & \colhead{Latitude Mean} & \colhead{Latitude FWHM} & \colhead{Latitude Peak}  \\
\colhead{(degrees)} & \colhead{(degrees)} & \colhead{(arcminutes)} & \colhead{(\# /degree$^2$)} \\
\colhead{(1)} & \colhead{(2)} & \colhead{(3)} & \colhead{(4)}}
\startdata
340-0  & $0.0\pm 0.1$   & 27 & 11.2 \\
  0-20 & $-0.2\pm 0.1$  & 24 & 11.2 \\
 20-40 & $0.03\pm 0.09$ & 28 & 10.8 \\
 40-60 & $0.0\pm 0.1$   & 52 &  3.4 \\
 60-80 & $0.0\pm 0.1$   & 53 &  2.0 \\
\enddata
\end{deluxetable}
%\clearpage

\begin{figure}
\plotone{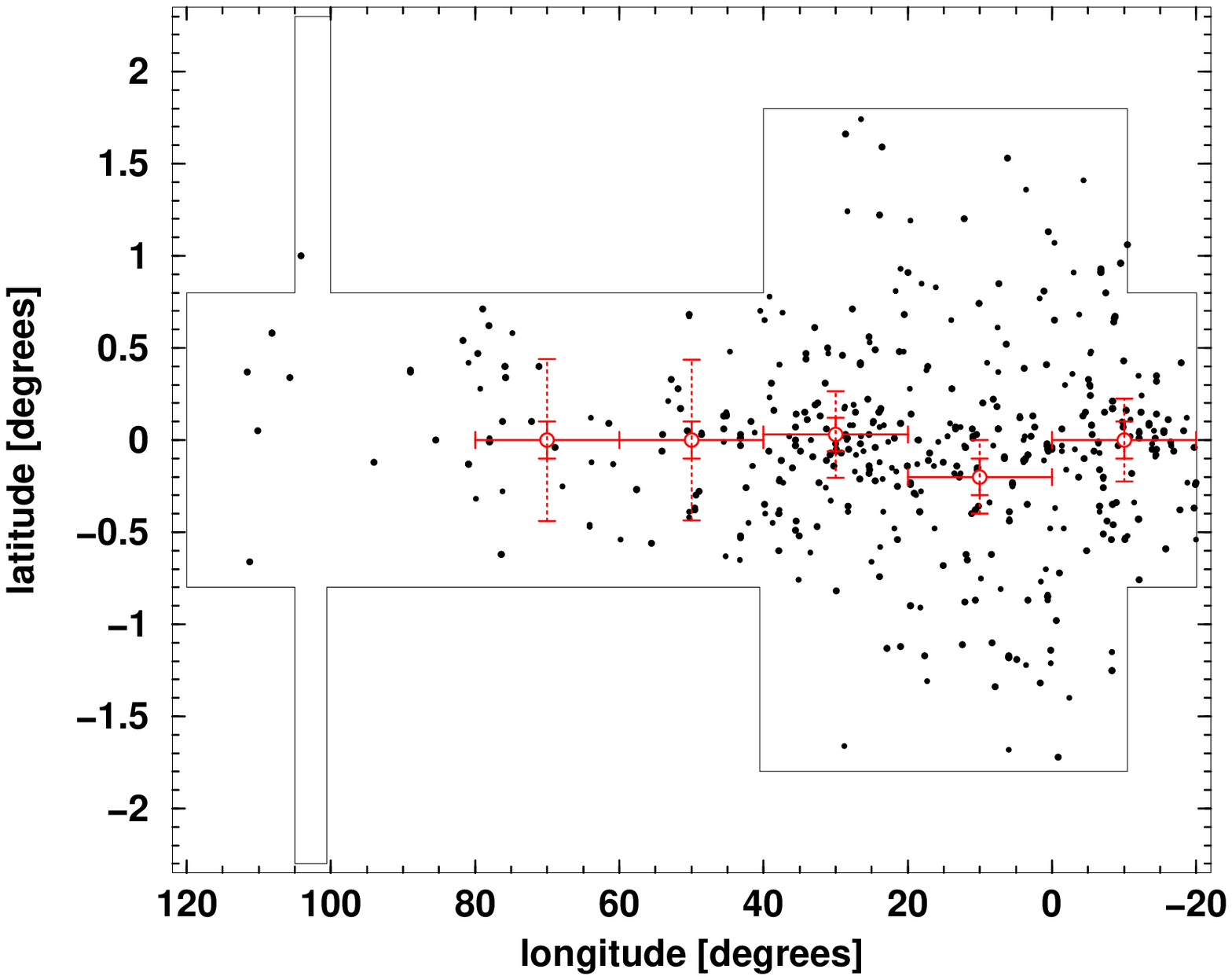}
\caption{Two-dimensional distribution of the matching subset with the boundaries of the survey overplotted. Average latitude values with their $1\sigma$ error bars (solid lines) and FWHM (dotted lines) are overplotted for longitude bins of $20^{\circ}$ (see Table \ref{bdist}). Notice the slight tilt in the concentration of sources along the Galactic plane between $-20^{\circ}$ and $20^{\circ}$. There is a significant drop-off in the number of sources and the width of the latitude distributions flattens out at longitudes greater than $\approx 40^{\circ}$.}
\label{coords_lb_20cm}
\end{figure}

These mean values are consistent with the known Galactic plane tilt with respect to $b=0^{\circ}$ (e.g., Hammersley et al. 1995), as is also illustrated in Figure \ref{coords_lb_20cm}, and was shown for the distribution of MSX6C-5 GHz matches in Paper I.

Figure \ref{coords_l_20cm} shows the longitude distribution for lower-latitude ($|b|< 0.4^{\circ}$) and for higher-latitude ($0.4^{\circ}<|b|<0.8^{\circ}$) sources, for which the coverage map is nearly uniform. For lower-latitude sources, the number of matching sources drops off significantly at $l\approx 40^{\circ}$ and almost completely vanishes $l\gtorder 80^{\circ}$, while for higher-latitude sources, the longitude distribution is much flatter.

%\clearpage
Figure \ref{coords_lb_20cm} shows the two-dimensional distribution.
%\clearpage

The overall spatial picture is that the population of matches drops off significantly at $l\approx 40^{\circ}$, and $b\approx \pm 0.4^{\circ}$. At $l\gtorder 40^{\circ}$ the width of the latitude distribution merely reflects the width of the coverage map. This behavior implies that the matching sample is dominated by population I sources. Since the infrared colors of these sources imply that they are embedded sources, we argue that they are predominantly young compact \hii regions. There is a latitude-independent component in the distribution, which can be attributed to the PNe population and/or extragalactic contamination. These populations should have a uniform latitude distribution in the survey area. We estimate their contribution to be $\ltorder 25\%$ of the sample - consistent with our estimate from \S \ref{rad_ir}.

\subsection{Morphology}
\label{morphology}

We use the average of the deconvolved major and minor axes of the 1.4 GHz sources as a measure of their sizes. The FWHM of the synthesized beam in the VLA B configuration is typically $5''$. The north-south elongation increases with decreasing longitude, and for sources at the southern declination limit it reaches $15''$. We compare the size distribution of the 1.4 GHz sources with MSX6C matches to the distribution of non-MSX6C sources. On average, the infrared-detected sources have larger sizes: $9.2''\pm 0.3''$ (median of 7.1'') compared to $3.3''\pm 0.1''$ (median of $2.0''$). The two subsets also show different average major axis to minor axis ratios: $2.1\pm 0.1$ (median of 1.7) and $3.3\pm 0.2$ (median of 2.2), respectively. However on average, the non-MSX6C subset is fainter at 1.4 GHz: the matching subset have average peak flux density $60\pm 3$ mJy (median of 29 mJy) and the non-MSX6C have average of $27.9\pm 0.8$ mJy (median of 11.0 mJy).
 This likely arises from a larger extragalactic component for the non-MSX subset which we would expect to be dominated by faint point sources along with some radio doubles (elongated sources with high major-to-minor-axis ratios) at the brighter end. There is also a significant contribution at the faint end from side-lobes. While the matching with the MSX6C catalog filters out these sources, the non-MSX subset may suffer from this effect.

\begin{figure}
\plotone{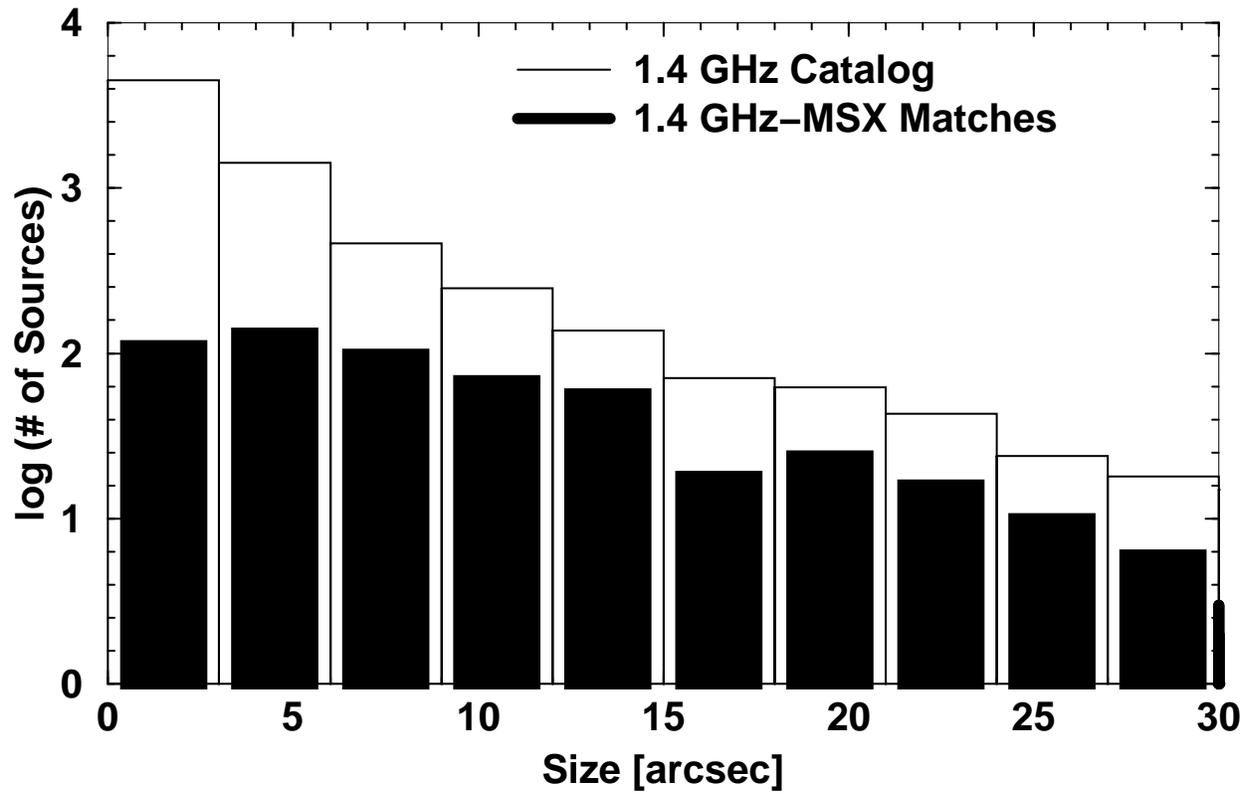}
\caption{Histogram of 1.4 GHz sizes for the entire radio catalog (empty bars), and for the MSX6C-1.4 GHz matching subset (filled bars).}
\label{sizes}
\end{figure}

In Figure \ref{sizes} we show a histograms of the size distribution of our 1.4 GHz sources and of the subset of MSX6C-1.4 GHz matches.

%\clearpage
%\clearpage

The first bin (size $\le 3''$) consists of unresolved sources. This plot shows clearly that extended 1.4 GHz sources are more likely to have MSX6C counterparts.

\begin{figure}
\plotone{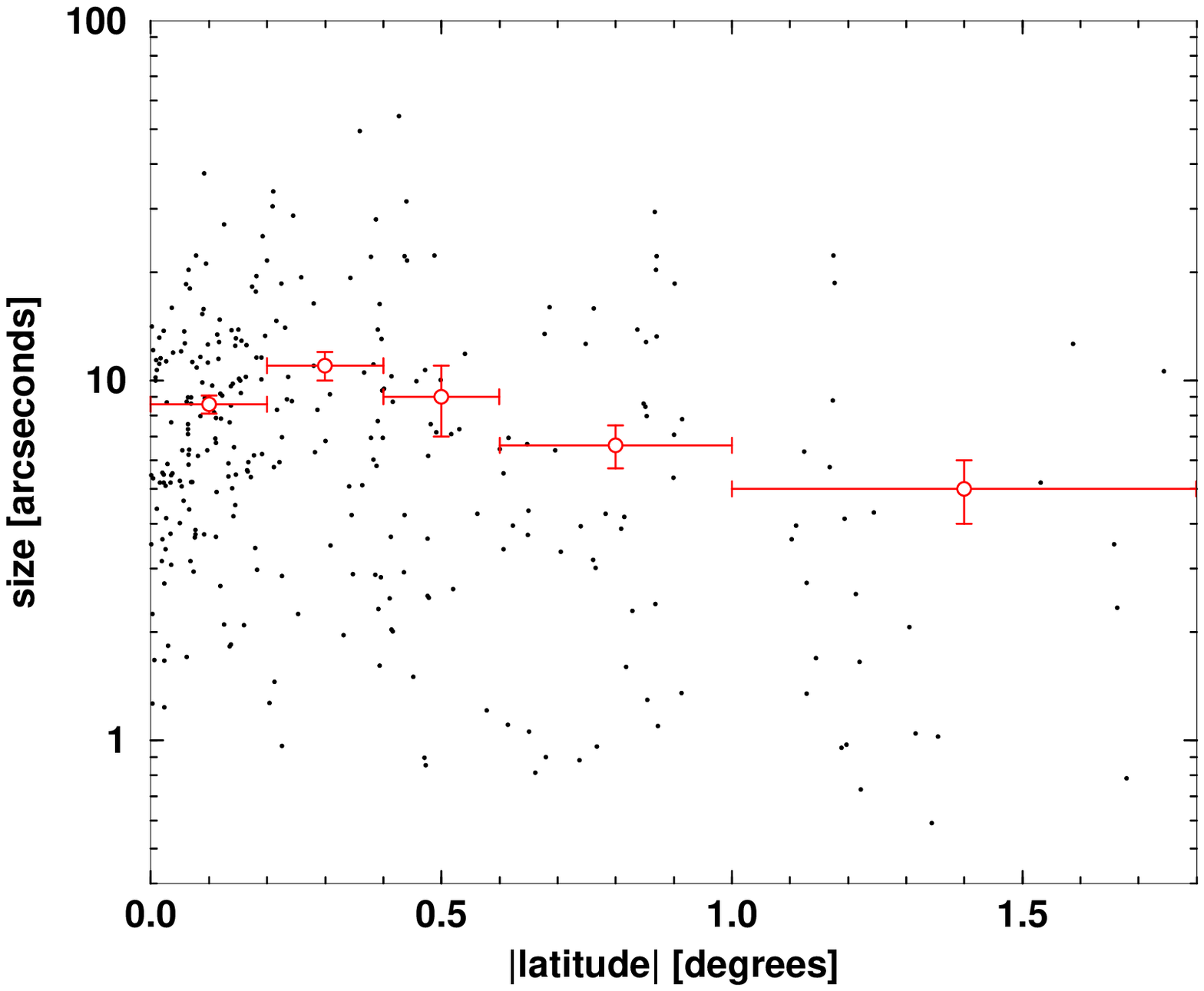}
\caption{The 1.4 GHz sizes as a function of the absolute values of latitude. Average size values with their $1\sigma$ error bars are overplotted for a few latitude bins.}
\label{bsizes}
\end{figure}

The structure of the 1.4 GHz coverage map allows us to check whether sizes depend on latitude. Figure \ref{bsizes} shows the 1.4 GHz sizes as a function of the absolute values of latitude. Higher-latitude sources have significantly smaller sizes.

%\clearpage
%\clearpage

Lower and higher-latitude sources do not have significantly different average major axis to minor axis ratios: $2.0\pm 0.1$ (median of 1.6) and $2.2\pm 0.1$ (median of 1.8), respectively. However on average, lower-latitude sources are brighter at 1.4 GHz, with average peak flux density of $65\pm 5$ mJy (median of 33 mJy) compared to $48\pm 5$ mJy (median of 24 mJy) for the higher-latitude sources. The infrared flux densities of the two latitude bins are consistent with each other.

\section{Conclusions}
\label{conc}

We have presented results from matching 1.4 GHz and MSX6C source catalogs of the Galaxy's first and second quadrants. The subset of matches extracted supplements the 5 GHz catalog of UC HII regions, which covers only the first Galactic quadrant, to give the most sensitive, high-resolution catalog of its kind to date. The completeness for detecting embedded O stars at 1.4 GHz is $\ge 50\%-90\%$, depending on the optical depth of the nebulae to free-free emission. We have investigated the infrared colors, the spatial distributions, and the morphologies of sources in this catalog. Our main findings are:
\begin{itemize}
\item[1.] The subset of 1.4 GHz sources with MSX6C colors is tightly confined to the Galactic plane (FWHM of $24-27'$ or $\sim 60-70$ pc). This suggests that the sample is dominated by Population I objects.
\item[2.] The number of MSX6C-1.4 GHz matches drops off sharply and the latitude distribution flattens out significantly at $l\gtorder 40^{\circ}$.
\item[3.] By matching our 1.4 GHz radio catalog and the MSX6C catalog, most of the blue MSX population is filtered out; the majority of the matching subset belongs to the red MSX population implying embedded sources. Our 1.4 GHz radio catalog and MSX6C infrared data cannot distinguish \hii regions from PNe.
\end{itemize}

The comprehensive compilation from the literature by Paladini et al. (2003) lists 765 known HII regions in our survey area, only 90 of which are smaller than $1'$ in diameter (at 2.7 GHz); thus, the matching subset described in this paper represents a 6-fold increase in the number of compact and ultra-compact HII regions in this region.

\section*{Acknowledgments}
This work was supported in part (R.H.B.) under the auspices of the US Department of Energy by Lawrence Livermore National Laboratory under contract W-7405-ENG-48, and the National Science Foundation under grant AST 0206309.
D.J.H. acknowledges the support of the National Science Foundation under grant AST 02-6-55.

\end{document}